\documentclass[twocolumn,showpacs,preprintnumbers,amsmath,amssymb,prl]{revtex4}

\usepackage{graphicx}
\usepackage{dcolumn}
\usepackage{bm}
\usepackage{epsfig}

\begin{document}

\title{Role of disorder and strong 5$d$ electron correlation in the electronic structure of Sr$_{2}$TiIrO$_{6}$}
\author{B. H. Reddy, Asif Ali, Ravi Shankar Singh}
\email{rssingh@iiserb.ac.in}
\affiliation{Department of Physics, Indian Institute of Science Education and Research Bhopal, Bhopal Bypass Road, Bhauri, Bhopal 462066 India}

\begin{abstract}

Transport and magnetic properties along with high resolution valence band photoemission study of disordered double perovskite Sr$_{2}$TiIrO$_{6}$ has been investigated. Insulator to insulator transition along with a magnetic transition concurrently occurs at 240 K. Comparison of valence band photoemission with band structure calculations suggests that the spin orbit coupling as well as electron correlation are necessary to capture the line shape and width of the Ir 5$d$ band. Room temperature valence band photoemission spectra show negligibly small intensity at Fermi energy, $E_{F}$. Fermi cut-off is observed at low temperatures employing high resolution. The spectral density of states at room temperature exhibits $|E-E_{F}|^{2}$ energy dependence signifying the role of electron-electron interaction. This energy dependence changes to $|E-E_{F}|^{3/2}$ below the magnetic transition evidencing the role of electron-magnon coupling in magnetically ordered state. The evolution of pseudogap ($\pm$12 meV) explains the sudden increase in resistivity ($\rho$) below 50 K in this disordered system. The temperature dependent spectral density of states at $E_{F}$ exhibiting $T^{1/2}$ behaviour verifies Altshuler-Aronov theory for correlated disordered systems.

\end{abstract}


\maketitle

Mott Hubbard model, consisting of large on-site Coulomb repulsion $U$ leading to strongly correlated narrow $d$ bands, has been very successful in explaining the electrical and magnetic properties of 3$d$ transition metal oxides (TMO) exhibiting noble physical phenomena such as metal insulator transition, colossal magneto-resistance, high-$T_{C}$ superconductivity, quantum criticality $etc$ \cite{mott, hubbard,rmp}. Since, $U$ depends on the extension of the radial wave function of the $d$ orbitals, 4$d$ and 5$d$ TMOs are expected to be weakly correlated wide band systems \cite{rmp,kalo1,kalo2}. Surprisingly, the importance of electronic correlation have been realised in various 4$d$ and 5$d$ systems \cite{nakatsuji, ravi1, ahn} exhibiting bad metallic to Mott insulating ground states and were interpreted as to have moderate to large $U$ leading the systems to be near the Mott criteria $U/W \sim $1 ($W$ = energy width of $d$ band) \cite{nakatsuji}. Apart from $U$, large spin orbit coupling (SOC) due to the larger ionic mass of 5$d$ ions also plays role to decide the ground state properties of 5$d$ TMOs. Among these, Iridium based perovskite related oxides are of special interest due to the fact that $U$ and SOC are of comparable energy scale. The cooperation between $U$ and SOC leads to form novel energy states designated by $J_{\mathrm{eff}}$, the effective total angular momentum state. Sr$_{2}$IrO$_{4}$ has been proposed as spin-orbit coupled Mott-insulator \cite{bjkim1}, in which the strong SOC lifts the orbital degeneracy of the $t_{2g}$ bands and results in such a narrow half-filled $J_{\mathrm{eff}}$ =$1/2$ band that even moderate electron correlation opens up a Mott gap in this 5$d$ electron system, which explains the insulating ground state. 

Along with $U$ and SOC, disorder may also play an important role in these wide 5$d$ band systems which can lead to Anderson type insulating behaviour due to the localization of carriers in the presence of strong disorder \cite{disorder}. Altshuler-Aronov theory for correlated disordered systems predicts that the density of states (DOS) in the vicinity of $E_{F}$ shows a singularity of $(E-E_{F})^{1/2}$ exhibiting a sharp dip while DOS at $E_{F}$ scales as $a+b\sqrt T$ \cite{AA, ddsarma}. Double perovskites have been widely studied due to the incorporation of two transition metals exhibiting wide variety of magnetic and electronic properties. Double perovskites can have ordered structure (rock salt or planer) having alternate arrangements (in 3 dimension or 2 dimension) or disordered structure having random arrangement of two different transition metals \cite {DP}. 5$d$ based disordered double perovskites give an opportunity to study the role of electron correlation, spin orbit coupling as well as disorder in the electronic structure of these systems. In particular, Sr$_{2}$TiIrO$_{6}$ forms in pseudo-cubic disordered double perovskite structure while the magnetic and transport measurements exhibit weakly ferromagnetic insulating ground state \cite {stio-bulk}.

In this letter, we investigate the electronic structure of Sr$_{2}$TiIrO$_{6}$ using photoemission spectroscopy to show that the strong correlation as well as SOC is important to capture the line shape and width of the Ir 5$d$ band. High resolution enables us to observe a Fermi cut-off in sharp contrast to the insulating transport. Energy dependence of the spectral DOS reveals the role of electron-magnon coupling in the magnetically ordered state. Evolution of sharp dip at $E_{F}$ explains the sudden increase of the resistivity below 50 K. Square root temperature dependence of the spectral DOS at $E_{F}$ verifies the Altshuler Aronov theory for correlated disordered systems.

Polycrystalline samples of Sr$_{2}$TiIrO$_{6}$ were prepared by conventional solid-state reaction route using high purity ingredients. Well ground powders were calcined at 600 $^o$C for 12 hours and finally sintered at 1200 $^o$C for 96 hours in pellet form. Absence of any secondary phases in x-ray diffraction measurements and longer sintering time leading to hard pellets ensures high quality of the samples. Magnetic and transport measurements were performed using Quantum Design SQUID-VSM and PPMS respectively. Photoemission spectroscopy were performed using Scienta R4000 electron energy analyser on \emph{in-situ} (base pressure $\sim$ 5x10$^{-11}$ mbar) fractured samples. Multiple samples were fractured and the reproducibility of the data were insured. Polycrystalline silver at 15 K was used to determine $E_{F}$ and the energy resolutions for different photon energies. Total instrumental resolutions were set to 400 meV, 8 meV and 5 meV for measurements with monochromatic Al \emph{K$_\alpha$} (1486.6 eV), He~{\footnotesize II} (40.8 eV) and He~{\footnotesize I} (21.2 eV) photons (energy) respectively.

\begin{figure}[tb]
\centering
\includegraphics[width=.58\textwidth,natwidth=650,natheight=750]{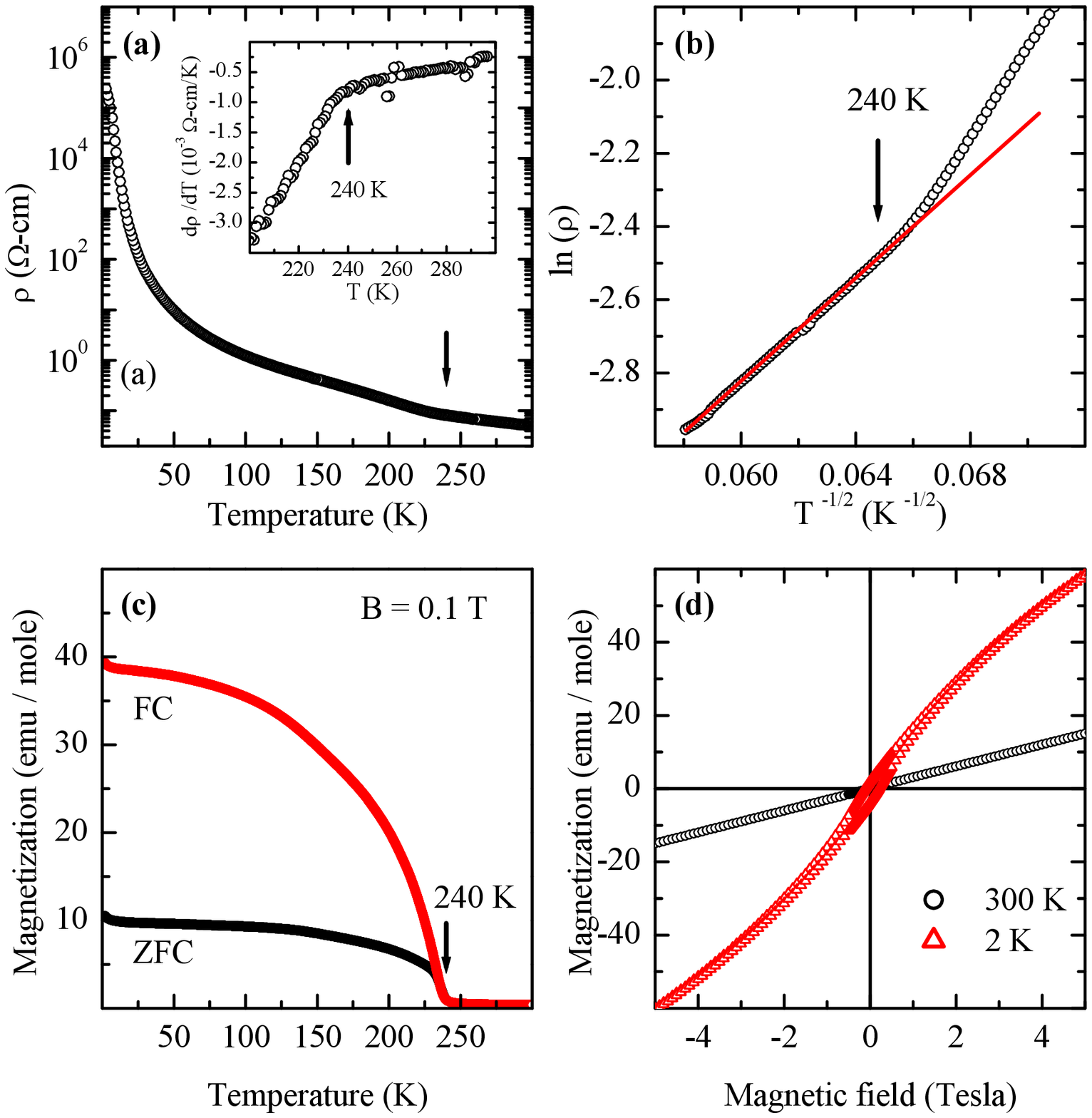}
\vspace{-32ex}
\caption{\label{fig:epsart} (color online) (a) Temperature dependent resistivity ($\rho$) in semi-log scale. Inset shows the first derivative of resistivity. (b) $\ln \rho$ vs $T^{-1/2}$ plot with linear fit exhibiting VRH mechanism at high temperature. (c) Temperature dependent ZFC and FC magnetization at 0.1 Tesla applied field. (d) Isothermal magnetization curves at 300 K and 2 K. Insulator to insulator and magnetic transitions at 240 K are marked by arrows.}
\end{figure}

Fig. 1(a) shows the temperature dependent resistivity ($\rho$) of Sr$_{2}$TiIrO$_{6}$. The slope of resistivity versus temperature curve is negative in the entire temperature range indicating insulating behaviour. The room temperature resistivity is about 0.05 $\Omega$-cm which increases to 10~$\Omega$-cm with decreasing temperature up to 50~K. The resistivity increases drastically below 50~K and reaches $\sim$ $10^{6}~\Omega$-cm at lower temperatures. An insulator to insulator transition at 240~K is marked by a down arrow which is clearer in the inset of the Fig. 1(a) where the change in the slop of $d\rho/dT$ is clearly evident at 240 K. In Fig. 1(b), we plot $\ln \rho$ vs $T^{-1/2}$ where the linear fit has also been shown between 300~K and 240~K. Temperature dependent $\rho$ exhibits Variable Range Hopping (VRH) of the form $\rho~\propto~\rho_{0}~$exp$(T_{0}/T)^{\nu}$ with $\nu$=$1/2$, where $\nu$= $(m+1)/(m+1+D)$ and $D$ is dimension of the system. The value of $\nu$=$1/2$ to be associated with the 3-dimensional VRH of carriers between states localized by disorder in the presence of electron correlation where DOS follows a power as $|(E-E_{F})|^{m}$ with $m$ = 2 \cite{vrh}. Fig. 1(c) shows the temperature dependent zero field cooled (ZFC) and field cooled (FC) magnetization curves obtained with 0.1 Tesla applied field. A magnetic transition is evident at 240 K below which the FC and ZFC curves bifurcate. Fig. 1(d) shows the isothermal magnetization curves taken at 300~K and at 2~K. Linear paramagnetic response at 300~K changes to opening of small hysteresis loop at 2~K suggesting weak ferromagnetism or canted antiferromagnetic ground state in this system as reported earlier \cite {stio-bulk}. The transport and magnetic measurements suggest that weakly insulating to highly insulating transition and a magnetic transition concurrently occurs at 240 K.

\begin{figure}[tb]
\centering
\includegraphics[width=.51\textwidth,natwidth=650,natheight=750]{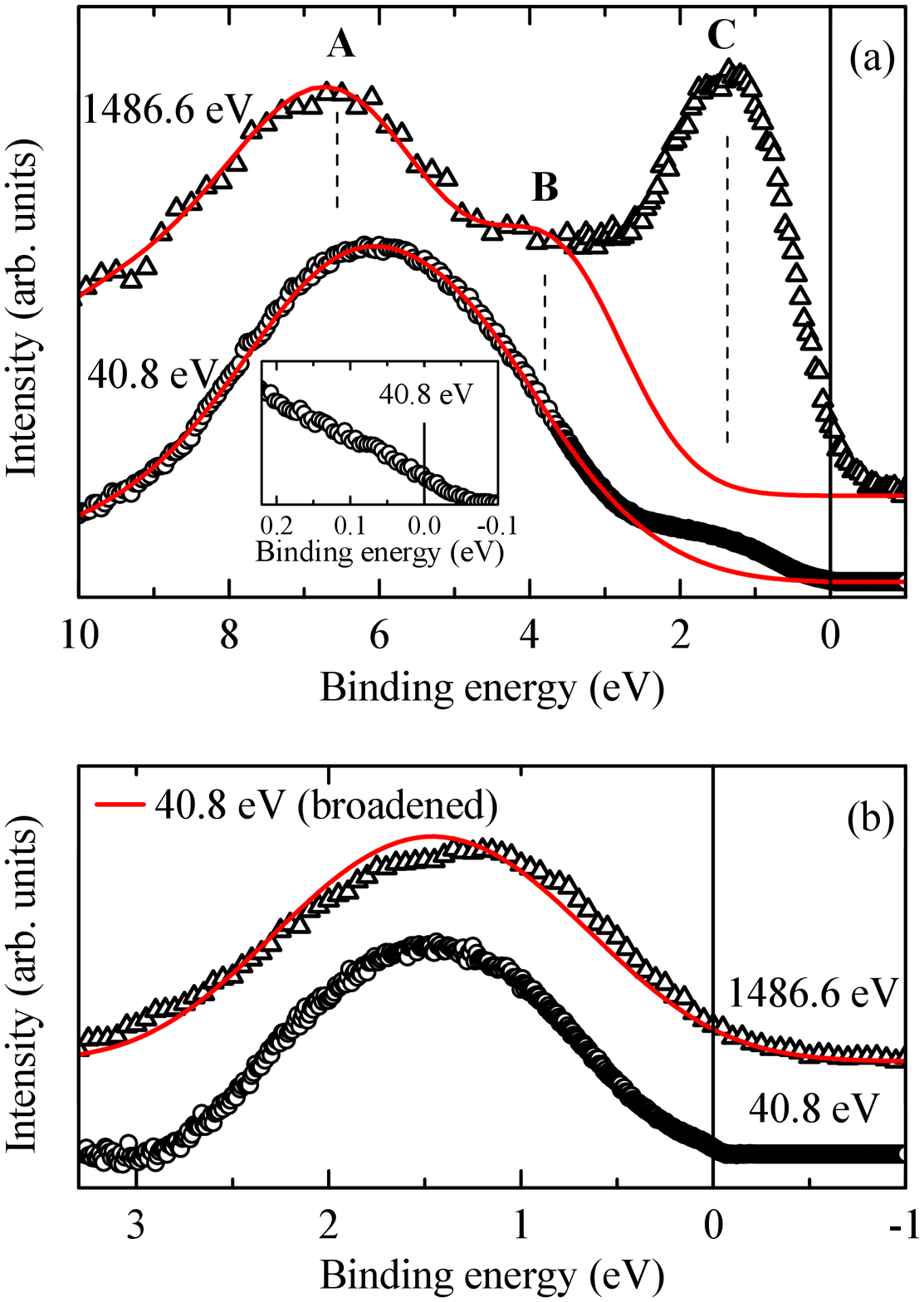}
\vspace{-17ex}
\caption{\label{fig:epsart} (color online) (a) XP and He~{\footnotesize II} valence band spectra. Lines show the O 2$p$ band contribution. Inset shows near $E_{F}$ region of He~{\footnotesize II} spectra. (b) Ir 5$d$ band extracted from XP and He~{\footnotesize II} spectra. Resolution broadened He~{\footnotesize II} is shown by lines.}
\end{figure}

Valence band spectra at room temperature obtained using 1486.6~eV (XP spectra) and 40.8~eV (He~{\footnotesize II} spectra) excitation energies are shown in Fig. 2(a). The valence band is formed by the hybridization of O 2$p$ with Ir 5$d$ states only, since, Sr and Ti states do not contribute in this energy range. Three discernible features A, B and C are marked by dashed vertical lines. Considering larger photoemission cross-section ratio of Ir 5$d$ states with O 2$p$ states for 1486.6~eV photons than for 40.8~eV photons \cite {yeh}, feature C appearing below 3~eV binding energy can be attributed to anti-bonding states having primarily Ir 5$d$ character while features A and B correspond to bonding and non-bonding states respectively having primarily O 2$p$ character. XP spectra shows finite intensity while He~{\footnotesize II} spectra exhibits negligibly small intensity at $E_{F}$ presumably due to larger resolution broadening in case of XP spectra. Inset shows the monotonous decrease of He~{\footnotesize II} spectral intensity in the vicinity of $E_{F}$. The feature C is distinctly separated from features A and B in both the spectra. Thus Ir 5$d$ band can reliably be delineated by subtracting the O 2$p$ bands using two gaussians corresponding features A and B as performed in other systems \cite{kalo1,kalo2,ravi1}. The resultant fit is shown by line in Fig. 2(a) and the extracted Ir 5$d$ contribution is shown in the Fig. 2(b). We also show the resolution broadened Ir 5$d$ band corresponding to He~{\footnotesize II} spectra in Fig. 2(b) by lines. The spectral line shape is very similar despite having large difference in probing depth in these two spectra suggesting that the surface and bulk electronic structures are very similar in this system. 

To address the role of $U$ and SOC on the electronic structure we performed band structure calculations for nonmagnetic ground state using full potential linearized augmented plane wave method (WIEN2k software) \cite{wien2k} within the local density approximations (LDA) for ordered Sr$_{2}$TiIrO$_{6}$ structure with experimental lattice constants. The convergence for different calculations were achieved considering 2000 $k$ points within the first Brillouin zone. The energy and charge convergence were set to 0.1 meV and 10$^{-4}$ electronic charge per formula unit (fu) respectively. LDA results have been shown in Fig. 3(a) showing total DOS as well as partial DOS corresponding to Ir and three crystallographically different O atoms. As seen in Fig. 2(a), bonding states (7.5 eV to 4.5 eV) and non-bonding states (4.5 eV to 1.5 eV) corresponding to features A and B respectively having primarily O 2$p$ character and anti-bonding states (1.5 eV to -0.25 eV) corresponding to feature C, having primarily Ir 5$d$ character are evident. An overall shift of about 1 eV to O 2$p$ band towards higher binding energy reproduces the experimental spectra. Such a shift in the completely filled O 2$p$ bands has often been observed due to large correlation present in the system \cite{ravi1, ddsarma1}. LDA calculations result in metallic ground state with Ir 5$d$ band width of about 1.75 eV centered around 0.5 eV binding energy which is far from experimentally observed width and peak position (Fig. 2(b)). To observe the evolution with inclusion of $U$ and/or SOC we show the results of LDA, LDA+$U$ ($U$=5 eV), LDA+SOC, and LDA+SOC+$U$ ($U$=5 eV) calculations in Fig. 3(b). Fermi function multiplied Ir 5$d$ PDOS is convoluted with the experimental resolution of 0.4 eV to compare with XP spectra. Interesting to note here that none of the calculations lead to opening of the gap at $E_{F}$, though there is a systematic change in the width and peak position of the 5$d$ band. Although other calculations \cite{stio-cal} suggest moderately large $U$ with SOC could open up the gap in these systems. LDA and LDA+$U$ both show very similar narrow 5$d$ band width centered around 0.5 eV while LDA+SOC results in widening of width such that the peak position shifts to little higher binding energy. It is only LDA+SOC+$U$ calculation which is somewhat closer to the experimental spectrum exhibiting larger band width which is centered above 1 eV binding energy. These observations suggest that inclusion of SOC as well as moderate to large $U$ is required to depict the experimentally observed width and finite intensity at $E_{F}$ in the photoemission spectra is well captured in LDA+SOC+$U$ calculations.

Interestingly, He~{\footnotesize II} spectra exhibits a Fermi cut-off (not shown here) when temperature is lowered to 15 K suggesting electronic states in the vicinity of $E_{F}$ corresponding to metallic ground state in contrast to insulating behaviour observed in transport measurements. These electrons at the $E_{F}$ being at the band edge with low Fermi velocity are possibly getting localised in the presence of strong intrinsic disorder in the system which leads to Anderson type insulating ground state \cite{disorder}.

\begin{figure}[tb]
\centering
\includegraphics[width=.57\textwidth,natwidth=650,natheight=750]{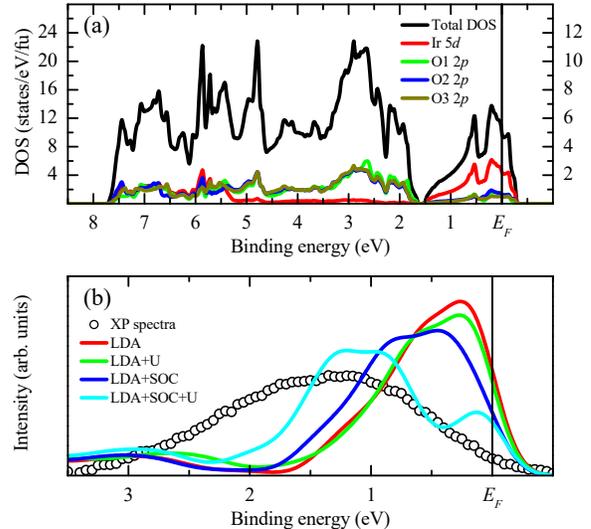}
\vspace{-36ex}
\caption{\label{fig:epsart} (color online) (a) Total DOS, Ir 5$d$ partial DOS, and three different O 2$p$ partial DOS obtained from LDA results. (b) Resolution broadened occupied Ir 5$d$ partial DOS corresponding to LDA, LDA+U, LDA+SOC and LDA+SOC+U. Symbols represent Ir 5$d$ band extracted from XP spectra.}
\end{figure}

\begin{figure}[tb]
\centering
\includegraphics[width=.57\textwidth,natwidth=650,natheight=750]{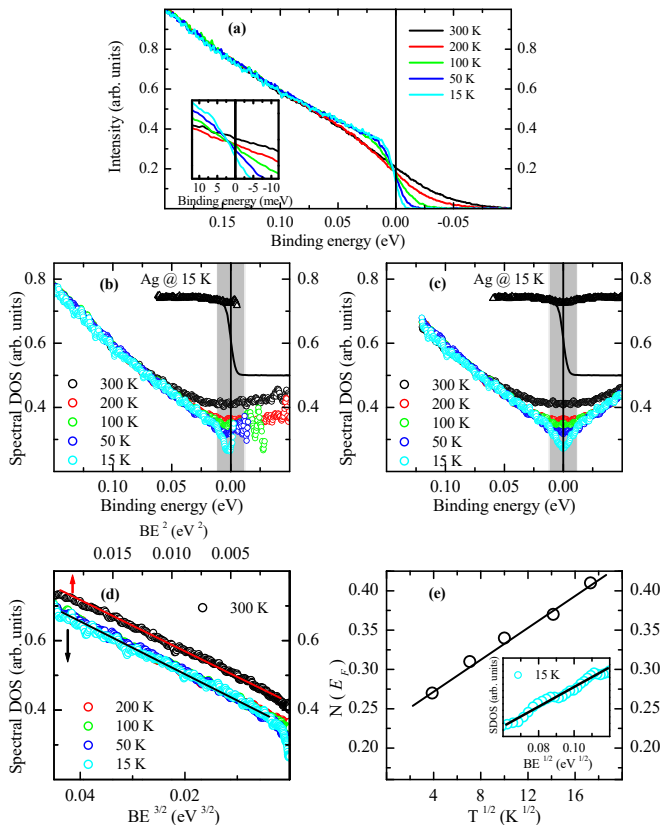}
\vspace{-6.5ex}
\caption{\label{fig:epsart} (color online) (a) High resolution He~{\footnotesize I} spectra. Inset shows the decreasing intensity at $E_{F}$ with decreasing temperature. Spectral DOS obtained (b) by dividing by resolution convoluted Fermi function and (c) by symmetrization. Vertically shifted Ag spectra and spectral DOS at 15 K are also shown with lines and symbols respectively. (d) Room temperature spectral DOS is plotted as a function of $(E-E_{F})^{2}$ (top scale) while lower temperature spectral DOS is plotted as a function of $(E-E_{F})^{3/2}$ (bottom scale). (d) Spectral DOS at $E_{F}$ verses $T^{1/2}$. Inset shows the $(E-E_{F})^{1/2}$ behaviour of spectral DOS at 15 K.}
\end{figure}

To further investigate the insulating behaviour and rapid increase of resistivity below 50 K, we show the temperature dependent high resolution spectra obtained using 21.2 eV (He~{\footnotesize I} spectra) excitation energy in Fig. 4(a). All the spectra corresponding to different temperatures normalised at 200 meV binding energy shows similar line shape down to about 75 meV. Interesting evolution of spectra near $E_{F}$ including appearance of a Fermi cut-off is evident as the temperature is lowered below 300~K. Spectral intensity at the $E_{F}$ gradually decrease with decreasing temperature as shown in the inset for better clarity. The decreasing intensity at the $E_{F}$ manifests the formation of pseudo gap as seen in other systems \cite{ddsarma}. 

Various physical and thermodynamic properties depend on the evolution of the electronic states within the thermal energy range ($k_{B}T$) around $E_{F}$. The high resolution employed in this study enables us to visualise the evolution of electronic states with temperature. The spectral DOS can be obtained by dividing the photoemission spectra by the resolution convoluted Fermi function since the electron and hole lifetime broadenings can be neglected in the vicinity of Fermi energy \cite {kalo2,ravi1}. To visualise the spectral evolution, we plot spectral DOS at different temperatures in Fig. 4(b). Vertically shifted Ag spectra as well as spectral DOS has also been shown in same figure. As expected, Ag exhibits flat spectral DOS in the vicinity of $E_{F}$. We also show the spectral DOS obtained by symmetrizing the photoemission spectra \cite {symm} in Fig. 4(c) along with the vertically shifted Ag spectra. Remarkable similarity in the spectral DOS obtained using these two different methods provides the confidence in the analysis procedure. 

Room temperature spectral DOS shows a parabolic energy dependence as shown in Fig. 3(d) by linear fit in spectral DOS versus $(E-E_{F})^{2}$ (top scale) plot. $(E-E_{F})^{2}$ energy dependence of DOS thereby exhibiting soft Coulomb gap \cite{vrh} at $E_{F}$ has also been seen in other correlated systems suggesting strong electron correlation among the Ir 5$d$ electrons \cite {ravi1}. As the temperature is reduced the spectral DOS at $E_{F}$ reduces significantly below the magnetic transition temperature (240 K) and starts to exhibit formation of pseudogap below 50 K which becomes a sharp dip like structure at the lowest measured temperature of 15 K. The width of the pseudo gap is about 24 meV as shown by the shaded grey region in the figure. All the spectra below 200 K show an energy dependence of $(E-E_{F})^{3/2}$ (bottom scale) as show in Fig. 3 (d) suggesting the influence of electron-magnon coupling on the electronic structure in the magnetically ordered state \cite{kalo2,ravi1,magnon}. Suggested by Altshuler-Aronov theory \cite{AA}, the sharp dip roughly following the $(E-E_{F})^{1/2}$ behaviour (within 12 meV to 5 meV range) is evident for the lowest temperature of 15 K as shown in the inset of Fig. 4 (e), while to verify the temperature dependence we plot the spectral DOS at $E_{F}$ (0.41, 0.37, 0.34, 0.31, and 0.27 at 300 K, 200 K, 100 K, 50 K, and 15 K respectively) versus $\sqrt T$ in Fig. 4 (e). A remarkable linear dependence, varifying Altshuler-Aronov theory, suggests the role of strong intrinsic disorder in the strongly correlated 5$d$ system.

In conclusion, we have investigated the electronic structure of disordered double perovskite Sr$_{2}$TiIrO$_{6}$. Surface and bulk electronic structures are very similar in this system. Observation of finite intensity at $E_{F}$ in room temperature spectra and its evolution to exhibite Fermi cut-off in lower temperature spectra is in contrast to the insulating transport suggesting Anderson insulating ground state. The larger width and line shape of Ir 5$d$ band suggests influence of spin orbit coupling and strong correlation among 5$d$ electrons. $(E-E_{F})^{3/2}$ dependence of spectral density of states in the high resolution spectra reveals the influence of electron magnon coupling in the magnetically ordered state. Appearance of sharp dip of width of about 24 meV explains the sudden increase of the resistivity below 50 K. The $\sqrt T$ dependence of the spectral DOS at $E_{F}$ verifies the Altshuler-Aronov theory for correlated disorder systems.

\end{document}